\documentclass[12pt]{article}

\catcode`\@=11

\global\arraycolsep=2pt
\usepackage{amsbsy,amssymb,latexsym,amsfonts,amsmath}
\usepackage{graphicx,color}

\begin{document}
\begin{flushright}
\parbox{4.2cm}
{CALT-TH 2015-007}
\end{flushright}

\vspace*{0.7cm}

\begin{center}
{\Large Local renormalization group functions from quantum renormalization group and holographic bulk locality}
\vspace*{1.5cm}\\
{Yu Nakayama}
\end{center}
\vspace*{1.0cm}
\begin{center}
{\it Walter Burke Institute for Theoretical Physics, California Institute of Technology,  \\ 
Pasadena, California 91125, USA}
\vspace{3.8cm}
\end{center}

\begin{abstract}
The bulk locality in the constructive holographic renormalization group requires miraculous cancellations among various local renormalization group functions. The cancellation is not only from the properties of the spectrum but from more detailed aspects of operator product expansions in relation to conformal anomaly.  It is remarkable that one-loop computation of the universal local renormalization group functions in the weakly coupled limit of the $\mathcal{N}=4$ super Yang-Mills theory fulfils the necessary condition for the cancellation in the strongly coupled limit in its $SL(2,\mathbf{Z})$ duality invariant form.
From the consistency between the quantum renormalization group and the holographic renormalization group, we determine some unexplored local renormalization group functions (e.g. diffusive term in the beta function for the gauge coupling constant) in the strongly coupled limit of the planar $\mathcal{N}=4$ super Yang-Mills theory.
\end{abstract}

\thispagestyle{empty} 

\setcounter{page}{0}

\newpage

The idea that the renormalization group scale may be regarded as the holographic direction has been appreciated since the birth of the AdS/CFT correspondence. In recent years, there have been various attempts to sharpen the idea so that we may derive the AdS/CFT from the first principle of the local (or quantum) renormalization group in the dual field theories \cite{Heemskerk:2010hk}\cite{Lee:2010ub}\cite{Lee:2012xba}\cite{Lee:2013dln}\cite{Mintun:2014gua}\cite{Lee:1215} It is, however, still mysterious what kind of cut-off and renormalization prescription should be used or how the bulk locality emerges in the strongly coupled regime. See also \cite{Behr} for a generalized viewpoint on the cut-off and the renormalization inspired by the bulk diffeomorphism.

Certainly, we do not expect that all quantum field theories should reveal the bulk locality in their holographic descriptions if any. There are various necessary conditions such as the factorization of correlation functions (``the large $N$ limit"), the sparseness of the low scaling dimension operators, the conditions on the central charges (or more generally those of OPEs) and so on (see e.g. \cite{ElShowk:2011ag} for summaries).
These conditions are typically deduced from what we empirically observe in the direct computations of the bulk gravity side, so it is instructive to see how (or even whether) these conditions show up from the first principle approach of the constructive holography alluded above. 

From the viewpoint of the constructive holography based on the local (quantum) renormalization group, it seems rather surprising that the higher derivative terms disappear in the bulk action. First of all, since there is no natural suppression parameter in the derivative expansions of the local renormalization group, there is no apparent reason why we can truncate the expansions at a certain derivative order.
If we could keep only the leading term in each expansions, we might say that such truncations would be attributed to (still mysterious) properties of strongly coupled gauge theories. However, the actual situation is much more involved \cite{Nakayama:2014cca}.  The constructive approach demands that certain higher derivative terms should necessarily appear as a consequence of the universal aspect of the local renormalization group, and what is actually happening is not the naive truncation in derivative expansions in each terms but miraculous cancellations among various seemingly unrelated terms in the local renormalization group evolution. 
We should emphasize that these local renormalization group functions are determined (in principle) by the OPE coefficients, so the cancellation is not only the properties of the spectrum of the CFT, but the dynamical information contained in the OPE structure.

In our recent paper \cite{Nakayama:2014cca}, we have discussed how such a cancellation mechanism is tightly related to the  
holographic condition of the central charges $a=c$, and how non-zero difference in central charges $a-c$ leads to the unique structure of the higher derivative corrections.
In a certain sense the AdS/CFT computation had predicted these results more than 15 years ago (because AdS/CFT works!), yet we managed to obtain new predictions for the structure of the local renormalizaiton group functions (such as the metric beta functions) as a byproduct of the consistency check.\footnote{Strictly speaking, it is unfortunate that our paper did not come out as ``a new prediction" because the purely AdS/CFT computation \cite{Jackson:2013eqa}\cite{Kiritsis:2014kua} (without reference to quantum renormalization group construction) appeared a couple of weeks earlier than ours while we were preparing for the final draft.}

In this paper, we investigate the cancellation mechanism further in the massless ``universal" sector of the AdS/CFT given by the metric (dual to the energy-momentum tensor) and the axion-dilaton (dual to the gauge coupling constant and theta angle ). As is clear, if we add more (massless) fields, the cancellation becomes more involved because the disparity in number becomes larger with more fields between the available local renormalizaiton group functions that can  be adjusted\footnote{In principle, these functions are determined once the theory is fixed so they are not even adjustable.} and the possible higher derivative terms that must be cancelled. 
Nevertheless, we do find that the cancellation does happen if we choose the non-universal local renormalization group functions properly.
Since the local renormalization group functions in  non-massless sector are generically not universally computable in a power-counting renormalization scheme, we may declare that our results essentially complete the consistency check of the bulk locality of the constructive holography in the $\mathcal{N}=4$ super Yang-Mills theory. While we will do present some plausible argument for the nature of the non-universal terms in our paper, we leave it for a future study to establish the precise Wilsonian cut-off scheme in which the non-universal local renormalization group functions can be unambiguously computed in agreement with our predictions.

Let us investigate the local renormalization group flow of the planar $\mathcal{N}=4$ super Yang-Mills theory in $d=4$ dimensions. We are particularly interested in the renormalization of the single trace energy-momentum tensor $T_{\mu\nu}$ that couples to the metric $g_{\mu\nu}(x)$ and the single trace ``holomoprhic Lagrangian" $L = \mathrm{Tr}\left(F_{\mu\nu}F^{\mu\nu} + iF_{\mu\nu} \tilde{F}^{\mu\nu} + \cdots \right)$ that couples to the (space-time dependent) ``holomorphic coupling constant" $\tau(x) = \frac{\theta(x)}{2\pi} + i\frac{4\pi}{g^2(x)}$ (and its multi trace composites). The main idea of the quantum renormalization group \cite{Lee:2010ub}\cite{Lee:2012xba}\cite{Lee:2013dln} is that we do not introduce the independent sources for the multi-trace operators out of $T_{\mu\nu}$  or $L$, but rather try to encode them in the bulk fields as a ``second quantization".

According to the prescription given in \cite{Lee:2010ub}\cite{Lee:2012xba}\cite{Lee:2013dln}, the Schwinger functional for the source $g_{\mu\nu}(x)$ and $\tau(x)$ will be expressed in terms of the bulk $d+1 = 5$ dimensional path-integral 
\begin{align}
e^{iW[g_{\mu\nu},\tau]} = \int \mathcal{D}X e^{i\int d^4x \sqrt{|g|} (\tau L(X) + h.c.)} = \int \frac{\mathcal{D} g_{\mu\nu}\mathcal{D}\pi^{\mu\nu} \mathcal{D}n \mathcal{D}n^\mu}{\mathrm{Diff}}\mathcal{D}\tau\mathcal{D}P  e^{iS} \ , \label{bulkpath}
\end{align}
where the bulk action in the Hamiltonian formulation takes the form
\begin{align}
S = \int dr d^4x  \left( \pi^{\mu\nu} \partial_r g_{\mu\nu} + P \partial_r \tau + \bar{P} \partial_r \bar{\tau} - n^\mu H_\mu - n H \right) \ .
\end{align}
The radial (dimensionless) coordinate $r$ is conventionally chosen as the renormalization group direction, and to make the radial coordinate and spatial coordinate $x^\mu$ on the same footing, it would be more convenient to introduce the scaling coordinate $z$ through $r = \log(z/z_0)$ such that $z$ has a dimension of length. 
Then by introducing the lapse field $\tilde{n} = n\frac{z_0}{z}$ and the shift field $\tilde{n}^\mu = n^\mu\frac{z_0}{z}$, we may reconstruct the $1+4$ dimensional metric $ds^2 = G_{MN} dx^M dx^N = \tilde{n}^2 dz^2 + g_{\mu\nu} (dx^\mu + \tilde{n}^\mu dz)(dx^\nu + \tilde{n}^\nu dz)$.
In addition, we identify $1+4$ dimensional dilaton-axion field as $\tau(x^\mu,z) = C + ie^{-\phi}$.  We will later fix the arbitrary dimensionful parameter $z_0$ from the requirement of the {\it manifest} $1+4$ dimensional diffeomorphism, but such a fixing is of course just conventional. Furthermore, in  order to simplify the following equations we choose the gauge in which the shift vector $n^\mu$  vanishes by using the scheme independence of the local renormalization group. We can always recover the shift dependence from imposing the $1+4$ dimensional diffeomorphism invariance (as long as the physics of the local renormalization group is scheme independent).

The Hamiltonian density $H$ is determined by the local renormalization group flow of the $\mathcal{N}=4$ super Yang-Mills theory: 
\begin{align}
H &= \sqrt{|g|}\Lambda[g_{\mu\nu},\tau] - \beta_{\mu\nu}[g_{\mu\nu},\tau] \pi^{\mu\nu} - \beta_{\mu\nu;\rho\sigma} [g_{\mu\nu},\tau]  \pi^{\mu\nu} \pi^{\rho\sigma} \cr
 &- \beta_\tau[g_{\mu\nu},\tau]  P - \bar{\beta}_{\bar{\tau}}[g_{\mu\nu},\tau] \bar{P} - \beta_{\tau \bar{\tau}}[g_{\mu\nu},\tau]  P\bar{P} \ . \label{action}
\end{align}
Here $\Lambda[g_{\mu\nu}, \tau]$ is the local renormalization of the ``cosmological constant" in the dual field theory.
$\beta_{\mu\nu}[g_{\mu\nu},\tau]  $ is the local renormalization group function (``beta function") of the $d=4$ dimensional metric $g_{\mu\nu}$ that couples to the single trace energy-momentum tensor $T_{\mu\nu}$, and  $\beta_{\mu\nu;\rho\sigma} [g_{\mu\nu},\tau] $ is the local renormalization group function for the double trace energy-momentum tensor  $:T_{\mu\nu} T_{\rho\sigma}:$. Similarly, $\beta_\tau[g_{\mu\nu},\tau]$ is the local renormalization group function of the local coupling constant $\tau(x)$ and $\beta_{\tau\bar{\tau}}[g_{\mu\nu},\tau] = \beta_{\bar{\tau}\tau}[g_{\mu\nu},\tau]$ is the local renormalization of the double trace operator.\footnote{It turns out that the double trace beta function has the K\"ahler structure.}

In order to determine the bulk action, we have to compute these local renormalization group functions in the (strongly coupled) $\mathcal{N}=4$ super Yang-Mills theory.  Since we should work in the Wilsonian local renormalization scheme, they contain higher derivative terms. As we have already emphasized, in general, it means that there is no reason to expect that the resulting bulk action is local. Indeed, we are going to see that each terms do contain higher derivative terms even in the strongly coupled limit, but they should eventually cancel out after integrating out the canonical momenta to obtain the Lagrangian formulation of the bulk action. This cancellation determines several renormalization group functions that are not computable within the power-counting renormalization scheme, but also requires the very specific structure of the local renormalization group functions that are computable within the power-counting renormalization scheme.

The power-counting renormalization scheme is tightly related to the removal of UV divergence in quantum field theories, and it computes the universal part of the more generic (Wilsonian) renormalization scheme. For instance, we may use the dimensional regularization, and subtract the power diverging part. The usual renormalizability argument shows that the higher dimensional operator does not mix with the lower dimensional operator in this scheme. 

In relation, we are quite implicit about the scheme choice. Even if we started with the scheme in which the higher derivative terms do not exsit for some reasons, we may always change the scheme by $g_{\mu\nu} \to \tilde{g}_{\mu\nu} (g_{\mu\nu}, \tau, R_{\mu\nu\rho\sigma}, \partial_\mu \tau \cdots)$ so that the renomalization group functions apparently contain higher derivative terms. From the $1+4$ dimensional viewpoint, this is nothing but the field redefinition of the bulk fields.\footnote{We will also fix the renormalizaiton scheme related to the $r$-direction reparametrization by choosing the conventional ``speed" of the local renormalization group transformation.} Our locality condition ``there is no higher derivative terms in the bulk" should be understood up to this field redefinition. In other words, when we require the cancellation among higher derivative terms, we are implicitly choosing such a good scheme.

Let us look at each local renormalization group functions. The computation of the local renormalization of the cosmological constant $\Lambda[g_{\mu\nu},\tau]$ in the $\mathcal{N}=4$ super Yang-Mills theory with $U(N)$ gauge group was recently completed by \cite{Buchbinder:2012uh} (as a generalization of the contribution  from the gauge fields obtained in \cite{Osborn:2003vk}) in one-loop perturbation theory. The universal part of their results show
\begin{align}
-\Lambda_{\mathrm{univ}}[g_{\mu\nu},\tau] &= \frac{N^2}{4(4\pi)^2} \left( 2\left(R_{\mu\nu} - \frac{1}{4}\frac{\partial_\mu \tau \partial_\nu \bar{\tau} + \partial_\nu \tau \partial_\mu \bar{\tau}}{(\mathrm{Im} \tau)^2} \right)^2 - \frac{2}{3}\left(R-\frac{\partial_\mu\tau \partial^\mu \bar{\tau}}{2(\mathrm{Im} \tau)^2}\right)^2 \right. \cr
 & \left. + \frac{1}{(\mathrm{Im} \tau)^2}\left|D^\mu \partial_\mu \tau + \frac{i}{\mathrm{Im} \tau} \partial_\mu \tau \partial^\mu \tau\right|^2  \right) \  , \label{1loop}
\end{align}
which is identified as the conformal anomaly with $a= c= \frac{N^2}{4(4\pi)^2}$.
It is noteworthy that this one-loop result is $SL(2,\mathbf{Z})$ invariant under $\tau \to \frac{a\tau +b}{c\tau + d}$.\footnote{For reader's convenience, we quote: $\mathrm{Im} \tau \to \frac{1}{(c\tau+d)(c\bar{\tau} +d)} \mathrm{Im} \tau$, $\partial_\mu \tau \to \frac{1}{(c\tau+d)^2}\partial_\mu \tau$ and $\left( D^\mu \partial_\mu \tau + \frac{i}{\mathrm{Im} \tau} \partial_\mu \tau \partial^\mu \tau\right) \to \frac{1}{(c\tau+d)^2} \left(D^\mu \partial_\mu \tau + \frac{i}{\mathrm{Im} \tau} \partial_\mu \tau \partial^\mu \tau \right) $ under $SL(2,\mathbf{Z})$ transformation.} This fact, however, is not so deep as it sounds. The trick is that at one-loop level, the conformal anomaly is just $N^2$ times that of the $U(1)$ theory which is obviously $SL(2,\mathbf{Z})$ invariant.
We will work in the renormalization group scheme in which $SL(2,\mathbf{Z})$ invariance of the $\mathcal{N}=4$ super Yang-Mills theory is manifest. It will be inherited to the $SL(2,\mathbf{Z})$ invariance of the dilaton-axion sector of the bulk type IIB gravity.

We should also realize that the  $SL(2,\mathbf{Z})$ invariant ``metric" $ds^2 = \frac{1}{(\mathrm{Im}\tau)^2} d\tau d\bar{\tau}$ that appears in front of the (generalized) Riegert four derivative kinetic operator\footnote{This operator, also known as Paneitz operator in some mathematics literature, first appeared in \cite{Fradkin:1982xc} in the context of superconformal gravity, and given our context, it may be most appropriate to name it after this paper, but we followed the convention.} (i.e. $(D^2)^2 +(2R_{\mu\nu}-\frac{2}{3}Rg_{\mu\nu}) D^\mu D^\nu$) in \eqref{1loop} is related to the Zamolodchikov metric  of the $\mathcal{N}=4$ super Yang-Mills theory. This is because the two-point functions of dimension four operators in $d=4$ dimensions is logarithmically divergent in flat space-time as $k^4\log k^2$ and this is the origin of the four derivative terms with quadratic on $\tau$ in the conformal anomaly. The Wess-Zumino consistency condition \cite{os} (see also \cite{Nakayama:2013is}\cite{Nakayama:2013wda}\cite{Jack:2013sha}\cite{Baume:2014rla}\cite{Auzzi:2015yia}) demands that at the conformal fixed point the flat space Laplacian squared $(\partial^\mu\partial_\mu)^2$ must be completed to be the Riegert operator.

Obviously, this computation provides the universal part determined from the power-counting renormalization scheme. In the Wilsonian local renormalization group scheme, we would add further lower dimensional terms
\begin{align}
\Lambda[g_{\mu\nu},\tau] = \Lambda_1[\tau] + \Lambda_{d\tau d\tau}[\tau] \frac{\partial^\mu \tau \partial_\mu \bar{\tau}}{(\mathrm{Im}\tau)^2}  + \Lambda_R[\tau] R + \Lambda_{\mathrm{univ}}[g_{\mu\nu},\tau] + \mathcal{O}(R^3) \ , \label{higher}
\end{align}
where $\Lambda_1[\tau]$, $\Lambda_{d\tau d\tau}[\tau]$, and $\Lambda_R[\tau]$ are some $SL(2,\mathbf{Z})$ invariant modular functions.
The power-counting renormalization scheme such as dimensional regularization does not say anything about these lower dimensional terms.
Note, however, that if we use the supersymmetric scheme, the potential term must not be renormalized so that it is natural to assume $\Lambda_1[\tau] = 0$, and since this is the only origin of the potential term in the bulk action when the single trace beta function vanishes for constant $\tau$, it is indeed agreement with the fact that the axion-dilaton does not have any potential term in the type IIB supergravity. Furthermore, within the same scheme, it was argued that the induced Newton constant $\Lambda_R[\tau]$ vanishes for the $\mathcal{N}=4$ super Yang-Mills theory at the one-loop order \cite{Fursaev:2000ym}\cite{Solodukhin:2015hma}.

It is worth noting that the condition that the Newton constant is not renormalized at one-loop is different from the other more familiar holographic prediction $a=c$. If we demand the both conditions for general massless field theories with weakly coupled Lagrangian descriptions, the matter contents must be multiples of $6$ real scalars, $4$ Majorana fermions and $1$ real vector. Indeed, the AdS/CFT with weakly coupled Lagrangian descriptions always have this matter contents (including $\mathcal{N}=4$ super Yang-Mills theory and its orbifold cousins).

 As a working hypothesis, we assume that the one-loop computation \eqref{1loop}  continues to be valid in the strongly coupled regime (see also \cite{Buchbinder:2012uh}\cite{Liu:1998bu}). 
We do not know of the formal proof of the non-renormalization theorem in particular for dilaton-axion interaction terms (e.g. $\partial_\mu\tau \partial^\mu \bar{\tau} \partial_\nu\tau\partial^\nu\bar{\tau}$ terms),
but we will eventually find that there is no room to modify it in order to reproduce what we know in the AdS/CFT  in the strongly coupled limit.
A possible argument for the non-renormalization theorem is that in the $\mathcal{N}=4$ superconformal field theories, the OPE of the energy-momentum tensor superconformal multiplet is completely fixed by one number given by the central charge $a=c$. The Wess-Zumino consistency condition of the conformal anomaly \cite{os}  (which amounts to the consistency of the Hamiltonian constraint in terms of quantum renormalization group) demands that this number does not depend on $\tau$ and all the quadratic terms in \eqref{1loop} are not renormalized. Since the OPE depends on the central charge alone, it is thus plausible that not only the quadratic term but also the entire conformal anomaly coming from the energy-momentum tensor superconformal multiplet are fixed by this number.
Presumably, this conformal anomaly is uniquely fixed from the $\mathcal{N}=4$ superconformal invariance with additional $SU(1,1)$ symmetry up to overall coefficient determined by the central charge $a=c$, but this has not been demonstrated in the literature.\footnote{Rather, this field theoretic one-loop computation has been the only way to fix the $SU(1,1)$ invariant $\mathcal{N}=4$ superconformal gravity so far known \cite{Buchbinder:2012uh}. Note, however, $SU(1,1)$ invariance is stronger than $SL(2,\mathbf{Z})$ invariance we could demand.} We will come back to this point at the end of this paper, but let us proceed under the working hypothesis for now. 

We move on to the double trace beta functions.
 In the strongly coupled regime, it is natural to assume that the derivative expansions of the local renormalization group functions  (up to a scheme choice). 
For the double trace beta functions, we have
\begin{align}
\beta_{\mu\nu;\rho\sigma} &= \frac{\eta(\tau)}{\sqrt{|g|}} \left(g_{\mu\rho} g_{\nu\sigma} -\lambda(\tau) g_{\mu\nu} g_{\rho\sigma} \right) + \mathcal{O}(D_\mu\tau, R) \cr
\beta_{\tau\bar{\tau}} &= \frac{\kappa(\tau)}{\sqrt{|g|}}(\mathrm{Im} \tau)^2 + \mathcal{O}(D_\mu\tau, R)  \ , \label{doubltrace}
\end{align}
where $\eta(\tau)$, $\lambda(\tau)$ and $\kappa(\tau)$ are $SL(2,\mathbf{Z})$ invariant modular functions.

First thing to notice here is that after integrating out the canonical momentum $\pi_{\mu\nu}$ and $P_{\tau}$, these double trace beta functions appear as the metric in the kinetic term of $\partial_z g_{\mu\nu}$ and $\partial_z\tau$ as $G^{\mu\nu;\rho\sigma} \partial_z g_{\mu\nu}\partial_z g_{\rho\sigma}$ and $\beta_{\tau\bar{\tau}}^{-1} \partial_z \tau \partial_z \bar{\tau}$, where $G^{\mu\nu;\rho\sigma}$ is the inverse of $\beta_{\mu\nu;\rho\sigma}$ (i.e. $\beta_{\mu\nu;\rho\sigma} G^{\rho\sigma;\eta\kappa} = \delta_{\mu}^\eta \delta_{\nu}^\kappa$).
Thus if we would like to obtain the second order action, there should be no higher derivative corrections in \eqref{doubltrace}. It also determines the modular weight of $\beta_{\tau\bar{\tau}}$ so that $\beta_{\tau\bar{\tau}}^{-1} \partial_z \tau \partial_z \bar{\tau}$ is invariant under $SL(2,\mathbf{Z})$ . 
There are further symmetry constraints. As discussed in \cite{Lee:2013dln} (see also \cite{Blas:2009yd}\cite{Henneaux:2009zb}), the consistency of the Hamiltonian constraint demands $\lambda = 1/3$. Of course, this is the value that we would get in the Einstein-Hilbert action. It is related to the fact that we did not have bare $R^2$ term in the conformal anomaly from the holographic renormalization group viewpoint \cite{Nakayama:2012sn}.

We now turn  to the single trace beta functions. One important consistency requirement in the holographic interpretation of the quantum renormalization group is that the single trace beta functions must be derived from the ``local boundary counter-term" as a gradient flow with respect to the double trace beta functions as their metric \cite{Lee:2012xba}\cite{Lee:2013dln}:
\begin{align}
\beta_{\mu\nu} &= \beta_{\mu\nu;\rho\sigma} \frac{ \delta S_B[g_{\mu\nu},\tau]}{\delta g_{\rho\sigma}} \cr
\beta_\tau &= \beta_{\tau \bar{\tau}} \frac{ \delta S_B[g_{\mu\nu},\tau]}{\delta \bar{\tau}} \ . \label{gradc}
\end{align}
Only when this gradient condition is satisfied, we may get rid of the first order $z$ derivative terms in the final bulk action as a boundary term after integrating out the canonical momenta.\footnote{This is not always necessary if the bulk action has first order in $r$ derivative such as in the case of topological coupling.}

We will expand the boundary counter-term as
\begin{align}
S_B = \int d^4x \sqrt{|g|} \left( -6\Lambda_B (\tau) - \kappa_B(\tau) R - \lambda_B(\tau) \frac{\partial_\mu \tau \partial^\mu \bar{\tau}}{(\mathrm{Im} \tau)^2} \right) \ ,
\end{align}
where $\Lambda_B(\tau)$, $\kappa_B(\tau)$ and $\lambda_B(\tau)$ are some modular functions invariant under $SL(2,\mathbf{Z})$.
From the double trace beta functions discussed just above and the gradient condition \eqref{gradc}, the single trace beta functions must be
\begin{align}
\beta_{\mu\nu} &= \eta(\tau)\left( \Lambda_B(\tau) g_{\mu\nu} + \kappa_B(\tau)\left( R_{\mu\nu}  - \frac{R}{6}g_{\mu\nu}  \right) \right. \cr 
& \left. + \lambda_B(\tau) \left(\frac{\partial_\mu\tau \partial_\nu \bar{\tau} + \partial_\mu \bar{\tau} \partial_\nu \tau}{2 (\mathrm{Im} \tau)^2} - \frac{g_{\mu\nu}}{6} \frac{\partial_\mu \tau \partial^\mu \bar{\tau}}{(\mathrm{Im} \tau)^2} \right)  + \frac{1}{2}\left(D_\mu \partial_\nu \kappa_B(\tau) + D_\nu \partial_\mu \kappa_B(\tau) \right)\right) \ . \cr
\beta_\tau &= (\mathrm{Im}\tau)^2 \kappa(\tau) \left(-6\partial_{\bar{\tau}} \Lambda_B(\tau) - \partial_{\bar{\tau}} \kappa_B(\tau) R\right) \cr
& + \kappa(\tau) \left(\lambda_B(\tau)\left(D^\mu \partial_\mu \tau + \frac{i}{\mathrm{Im} \tau} \partial_\mu \tau \partial^\mu \tau\right) + \partial_\mu \lambda_B(\tau) \partial^\mu\tau -(\partial_{\bar{\tau}}\lambda_B(\tau) ) \partial_\mu\tau \partial^\mu \bar{\tau}   \right) \ . \label{singletrace}
\end{align}
Let us fix some parameters from what we know in $\mathcal{N}=4$ super Yang-Mills theory.
Since the gauge coupling constant of the $\mathcal{N}=4$ super Yang-Mills theory does not run when $\tau = \mathrm{const}$, $\Lambda_B(\tau)$ must be independent of $\tau$. 
It is presumably also true that the beta function does not depend on the background curvature $R$, so $\kappa_B(\tau) = \kappa_B$, which is independent of $\tau$, is plausible, but we do not assume it for now, and rather we will derive it later from the requirement of the cancellation among beta functions.

Since we have introduced all the ingredients at this point, now we would like to demonstrate how the cancellation of various four derivative terms happen in the final bulk action. We first integrate out the canonical momenta to get the Lagrangian formulation of the bulk action
\begin{align}
S = \int dr d^4x  &\left(-n^{-1}G^{\mu\nu;\rho\sigma} \partial_r g_{\mu\nu} \partial_r g_{\rho\sigma} - n\frac{G^{\mu\nu;\rho\sigma}}{4} \beta_{\mu\nu} \beta_{\rho\sigma} \right.  \cr
& \left. - n^{-1}\beta_{\tau \bar{\tau}}^{-1} \partial_r \tau \partial_r \bar{\tau} - n \beta^{-1}_{\tau \bar{\tau}} \beta_\tau \beta_{\bar{\tau}} - n\sqrt{|g|} \Lambda[g_{\mu\nu},\tau] \right) \ .
\end{align} 
Substituting \eqref{doubltrace} and \eqref{singletrace}, we have the contribution to the four derivative terms from $\frac{G^{\mu\nu;\rho\sigma}}{4} \beta_{\mu\nu} \beta_{\rho\sigma} + \beta^{-1}_{\tau \bar{\tau}} \beta_\tau \beta_{\bar{\tau}}$ as well as from $\Lambda_{\mathrm{univ}}[g_{\mu\nu},\tau]$ contained in $\Lambda[g_{\mu\nu},\tau]$. It is interesting to observe that given the gradient nature, the potential term from $G^{\mu\nu;\rho\sigma} \beta_{\mu\nu} \beta_{\rho\sigma}$ gives the interaction that satisfies the detailed balance condition advocated in \cite{Horava:2008ih}\cite{Horava:2009uw}.\footnote{In \cite{Griffin:2011xs}\cite{Griffin:2012qx}, the further connection between detailed balance condition and conformal anomaly was discussed from the bulk viewpoint.} 
 In order to  get rid of all the four derivative terms in the bulk action, the terms that satisfy the detailed balanced condition must cancel against the conformal anomaly. Or in other words, the conformal anomaly must satisfy the detailed balance condition, which is by far the very non-trivial necessary condition for the AdS/CFT with the second order bulk action to work.

In addition to the condition $\Lambda_B(\tau) = \Lambda_B$ we have already discussed from the exactly marginal coupling constant of the $\mathcal{N}=4$ super Yang-Mills theory, the cancellation demands various relations among the beta functions such as 
\begin{align}
\kappa_B(\tau) & = -2\lambda_B(\tau) = \kappa_B \cr
\kappa(\tau) &= \frac{\eta(\tau)}{2} = \kappa \ . \label{fixed1}
\end{align}
which essentially remove all the possible non-trivial modular functions in relation to the instanton contributions to the local renormalization group functions beyond one-loop.\footnote{We should stress that these modular functions depend on $N$, and may be expressed in 't Hooft $1/N$ expansions while keeping $\lambda_{\mathrm{'t Hooft}} = g^2N$. 
It is logically possible that this cancellation occurs only in the strongly coupled regime $\lambda_{\mathrm{'t Hooft}} \to \infty$ so that we do not claim that this constancy is proved in the intermediate coupling even in the planer limit we consider.}
There are further normalization condition we would employ. The first is the overall normalization between the conformal anomaly and the four derivative terms giving $\kappa \kappa_B^2 = 4c$, where we recall $c=\frac{N^2}{4(4\pi)^2}$ is the central charge. Relatedly, in order to precisely connect the local renormalization group and the conformal anomaly, we had to provide the canonical normalization of the local renormalization group transformation. The computation that led to \cite{Buchbinder:2012uh}  should have corresponded to $\kappa\Lambda_B = 1$ so that the engineering Weyl weight of the $d=4$ dimensional metric is two (i.e. $\beta_{\mu\nu} = 2g_{\mu\nu} + \mathcal{O}(R)$) from \eqref{singletrace}.

In this way, we have determined the single trace as well as double trace beta functions of the $\mathcal{N}=4$ super Yang-Mills theory in the strongly coupled limit as
\begin{align}
\beta_{\mu\nu} &= 2g_{\mu\nu} + \frac{8c}{\kappa_B} \left(R_{\mu\nu} - \frac{R}{6} g_{\mu\nu} \right) - \frac{8c}{\kappa_B} \left( \frac{\partial_\mu\tau \partial_\nu \bar{\tau} + \partial_\mu \bar{\tau} \partial_\nu \tau}{4 (\mathrm{Im} \tau)^2} - \frac{g_{\mu\nu}}{12} \frac{\partial_\mu \tau \partial^\mu \bar{\tau}}{(\mathrm{Im} \tau)^2} \right) \cr
\beta_{\tau} & =  -\frac{2c}{\kappa_B} \left(D^\mu \partial_\mu \tau + \frac{i}{\mathrm{Im} \tau} \partial_\rho \tau \partial^\rho \tau\right) \cr
\beta_{\mu\nu;\rho\sigma} & = \frac{8c}{\kappa_B^2 \sqrt{|g|}} \left(g_{\mu\rho}g_{\nu\sigma} - \frac{1}{3} g_{\mu\nu}g_{\rho\sigma} \right) \cr
\beta_{\tau\bar{\tau}} & = \frac{4c}{\kappa_B^2 \sqrt{|g|}} (\mathrm{Im}\tau)^2  \ .  \label{fixed2}
\end{align}
In particular, there is no {\it dimensionless} arbitrary parameter left in the local renormalization group functions.

We have a couple of comments about the so-determined local renormalizaiton group functions. Firstly, we see that the local renormalization group flow of the metric and the coupling constant has a diffusive nature as observed in \cite{Jackson:2013eqa}. The purely curvature dependent part of the metric beta function given by the Schouten tensor in holography has been given in \cite{Verlinde:1999xm}\cite{Jackson:2013eqa} (up to the scheme choice elucidated in \cite{Kiritsis:2014kua}), and our results are in agreement. However, its microscopic origin is not so obvious. It appears that the leading derivative corrections  come from the zeroth order in coupling constant, but we do not know how this is obtained in the perturbative computation, and it may be very particular to the strongly coupled limit. 
Secondly, while the local renormalization group equation may suggest that the single trace beta functions show the gradient property in some situations (see e.g. \cite{os}), the metric used for the gradient condition there
 does not have to coincide with the double trace beta functions. Again this seems very peculiar to the models with holographic duals.

Once we determine these local renormalization group beta functions, we see that all the higher derivative terms are automatically cancelled, and the final bulk action is second order in derivative expansions.
We still do not know the precise details of $\Lambda_{d\tau d\tau}[\tau]$ and $\Lambda_R[\tau]$ in \eqref{higher} in the strongly coupled regime, but one-loop results quoted there suggests that they vanish. Independently if we would like to reproduce what we know in the AdS/CFT correspondence, they should vanish in the strongly coupled limit. In either viewpoint we take, assuming that they vanish, the final bulk action becomes 
\begin{align}
S &= \int dr d^4x \sqrt{|g|} n \frac{ \kappa_B^2}{8c} \left( 12 + \frac{4c}{\kappa_B}R - \frac{4c}{\kappa_B} \frac{\partial_\mu \tau \partial^\mu \bar{\tau}}{2(\mathrm{Im}\tau)^2} \right. \cr
 &  \left.  \ \ \ \ \ \ \ \ \ \ \ -  n^{-2}(g^{\mu\rho}g^{\rho\sigma} -g^{\mu\nu} g^{\rho\sigma}) \partial_r g_{\mu\nu} \partial_r g_{\rho\sigma} - 2n^{-2}\frac{\partial_r \tau \partial_r \bar{\tau}}{(\mathrm{Im} \tau)^2}  \right) \cr 
&= \int dz d^4x \sqrt{|g|} \tilde{n} \left( \frac{3\kappa_B^2}{2c z_0} + \frac{\kappa_B} {2 z_0}R - \frac{\kappa_B}{2 z_0} \frac{\partial_\mu \tau \partial^\mu}{2 (\mathrm{Im} \tau)^2} \right. \cr
& \left. \ \ \ \ \ \ -\tilde{n}^{-2} \frac{\kappa_B^2 z_0}{2c} (g^{\mu\rho}g^{\rho\sigma} -g^{\mu\nu} g^{\rho\sigma}) \frac{\partial_z g_{\mu\nu}}{2} \frac{\partial_z g_{\rho\sigma}}{2}  - \tilde{n}^{-2}\frac{\kappa_B^2 z_0}{4c}  \frac{\partial_z \tau \partial_z \bar{\tau}}{(\mathrm{Im} \tau)^2}  \right) .
\end{align}
Here, we recall  $r = \log(z/z_0)$ and $\tilde{n} =  n\frac{z_0}{z}$.
At this point, we may choose the arbitrary dimensionful parameter ${z}_0 = \sqrt{\frac{c}{\kappa_B}}$ such that the $(1+4)$ dimensional diffeomorphism is manifest.\footnote{We note that if  $\Lambda_{d\tau d\tau}[\tau]$ and $\Lambda_R[\tau]$ were not proportional to the combination appearing here, the $(1+4)$ dimensional diffeomorphism would be broken at this point. Furthermore, if they were non-zero, we would not recover the holographic conformal anomaly as we check below.} 
Then the resulting $(1+4)$  dimensional bulk action
\begin{align}
S = \int d^5x\sqrt{|G|} \left(\frac{3\kappa_B^{5/2}}{2c^{3/2}} + \frac{\kappa_B^{3/2}}{2c^{1/2}} \left(R_{(5)} -  \frac{\partial_M \tau \partial^M \tau}{2(\mathrm{Im}\tau)^2} \right) \right)
\end{align}
is nothing but the effective type IIB supergravity action on $AdS_5\times S^5$ restricted on the metric-dilaton-axion sectors albeit in a slightly unconventional normalization of dimensionful parameters. To see that the normalization is nevertheless correct, we may compute the holographic central charges \cite{Henningson:1998gx} from $a_{\mathrm{hol}} = c_{\mathrm{hol}} = \frac{1}{G_5\Lambda^{3/2}}$, where the normalization of the Einstein-Hilbert action is  $ S = \frac{1}{2G_5}\int d^5x\sqrt{|G|} \left( R_{(5)} + 3\Lambda \right)$. Substituting our number gives back to the $c$ we started with in the $\mathcal{N}=4$ super Yang-Mills theory as expected. In passing, we also note that in type IIB superstring theory, the higher derivative corrections start at $\mathcal{O}(R^4)$ and the difference between $N^2$ and $N^2-1$ in relation to $U(N)$ versus $SU(N)$ is even higher order (see e.g. \cite{Green:1999qt}).

The introduction of the other supergravity fields will not be so difficult.
The dual supergravity fields for the entire $\mathcal{N}=4$ energy-momentum tensor superconformal multiplet contains the additional $SO(6)$ adjoint valued vector fields, the anti-symmetric anti-self dual tensor in $SO(6)$ vector representation, the complex scalar in $10$ representation of $SO(6)$ and the real scalar in $20$ representation of $SO(6)$. The conformal anomaly for these fields have been also derived in \cite{Buchbinder:2012uh} at one-loop level. 
We have not discussed them because all these terms contain no higher derivative terms so we would not expect any non-trivial cancellation among various local renormalization group functions.
By using the same construction above, however, it is conceivable that we may construct the entire low energy effective bulk action (with some further ansatz on the single trace as well as double trace beta functions). 

For instance, it is trivial to see that the gauge kinetic term $\mathrm{Tr}F_{\mu\nu}F^{\mu\nu}$ of $SO(6)$ for the vector fields induced by the conformal anomaly (see e.g. \cite{Buchbinder:2012uh}) is precisely what we would get in the effective type IIB supergravity, which reproduces the correct current central charges from the holographic dictionary. One thing to be noticed, however, is that the gauging of $SO(6)$ is anomalous. Thus, the Schwinger functional for the $SO(6)$ background gauge fields is not gauge invariant. It is well-known (see e.g. \cite{Witten:1998qj}) that such non-gauge invariance is supplied by the $1+4$ dimensional Chern-Simons functional, and this is precisely what we need in the type IIB supergravity in agreement with the holography. Since the Chern-Simons functional is topological, there is no problem in identifying the extra direction appearing in the anomaly action with our holographic direction. We also note that the essential features of the local renormalization group analysis is not affected by the anomaly (see e.g. \cite{Baume:2014rla}\cite{Auzzi:2015yia} for detailed analysis).

Let us come back to the question of one-loop exactness of the conformal anomaly contribution to the local renormalization of the cosmological constant. As observed in \cite{Osborn:2003vk}, the bosonic contribution to the conformal anomaly in $\mathcal{N}=4$ super Yang-Mills theory shows the non-zero two-loop corrections to the interaction terms of 
\begin{align}
\delta \Lambda_{\mathrm{univ}} = \frac{\zeta(\tau)}{(\mathrm{Im} \tau)^4} (\partial^\mu \tau \partial_\mu \bar{\tau})^2 + \frac{\xi(\tau)}{(\mathrm{Im} \tau)^4} (\partial^\mu \tau \partial_\mu {\tau}\partial^\nu \bar{\tau}\partial_\nu \bar{\tau})   \ , \label{tloop}
\end{align}
and it was conjectured that these coefficients $\zeta(\tau)$ and $\xi(\tau)$ will be certain modular invariant functions.
However, as we have already seen in \eqref{fixed1}\eqref{fixed2}, we have fixed all the other local renormalization group functions to cancel the other one-loop exact terms of $\Lambda[g_{\mu\nu},\tau]$.
If the term like \eqref{tloop} are generated by the renormalization group flow and takes a different functional form in the strongly coupled regime, the AdS/CFT that we know today would not work. Therefore, we believe that the entire conformal anomaly is not renormalized beyond the one-loop order, and for this two-loop computation, we conjecture that the fermionic contribution precisely cancels against it.\footnote{Indeed, the contribution from vector fields studied in \cite{Osborn:2003vk} alone would not show the consistent results with AdS/CFT even at the one-loop level. The additional contribution from the fermions as well as scalars added in \cite{Buchbinder:2012uh} is imperative for the cancellation in order to make the bulk locality in AdS/CFT possible.} It would be very interesting to see if this is indeed the case or not by explicit two-loop computations and/or the constraint from $\mathcal{N}=4$ superconformal invariance.

\section*{Acknowledgements}
I would like to thank S.~S.~ Lee and S.~Solodukhin for the discussions and correspondence.
I also would like to thank the organizers of ``Holographic Renormalization Group and Entanglement" at APC Paris for the invitation and the stimulating atmosphere I enjoyed very much. 
This work is supported by Sherman Fairchild Senior Research Fellowship at California Institute of Technology  and DOE grant number DE-SC0011632.
After the appearance of the first version of the paper on the arxiv, we had fruitful discussions with P.~Horava, H.~Osborn, and A.~Tseytlin.

\end{document}